\PassOptionsToPackage{colorlinks=true,linkcolor=blue,urlcolor=blue,citecolor=blue}{hyperref}
\documentclass[fleqn,10pt]{wlscirep}
\usepackage{graphicx} 
\usepackage{float} 
\usepackage{dcolumn} 
\usepackage{bm,color}
\usepackage{amsmath,amssymb,dsfont,amstext,amsfonts}
\usepackage{extarrows}
\usepackage{xcolor}
\usepackage{wasysym}
\usepackage{mathtools}
\usepackage{bbold}
\usepackage[export]{adjustbox}
\usepackage{mathdots}
\usepackage{siunitx}
\usepackage[colorlinks=true,linkcolor=blue,urlcolor=blue,citecolor=blue]{hyperref}
\usepackage{mhchem}
\usepackage{mathrsfs}
\usepackage{multicol}
\usepackage[numbers,sort&compress]{natbib}

\usepackage[normalem]{ulem} 
\usepackage{color}

\usepackage[labelfont=bf]{caption}
\renewcommand{\aa}{\hat{a}^\dag}

\newcommand{\ie}{i.e.\ }

\title{Impact of electronic correlations on the superconductivity of high-pressure CeH$_9$}

\author[1,2,3,*]{Siyu Chen}
\affil[1]{TCM Group, Cavendish Laboratory, University of Cambridge, J. J. Thomson Avenue, Cambridge CB3 0HE, United Kingdom}
\affil[2]{Department of Materials Science and Metallurgy, University of Cambridge, 27 Charles Babbage Road, Cambridge CB3 0FS, United Kingdom}
\affil[3]{European Theoretical Spectroscopy Facility, Institute of Condensed Matter and Nanosciences, Université catholique de Louvain, Chemin des Étoiles 8, B-1348 Louvain-la-Neuve, Belgium}
\author[4]{Yao Wei}
\affil[4]{Department of Physics, King’s College London, Strand, London WC2R 2LS, United Kingdom}
\author[2]{Bartomeu Monserrat}
\author[4,5,*]{Jan M. Tomczak}
\affil[5]{Institute of Solid State Physics, TU Wien, 1040 Vienna, Austria}
\author[3,6,*]{Samuel Ponc\'e}
\affil[6]{WEL Research Institute, avenue Pasteur 6, 1300 Wavre, Belgium}
\affil[*]{e-mails: sc2090@cam.ac.uk, jan.tomczak@kcl.ac.uk, samuel.ponce@uclouvain.be}

\begin{abstract}
Rare-earth superhydrides have attracted considerable attention because of their high critical superconducting temperature under extreme pressures. 
They are known to have localized valence electrons, implying strong electronic correlations. 
However, such many-body effects are rarely included in first-principles studies of rare-earth superhydrides because of the complexity of their high-pressure phases. 
In this work, we use a combined density functional theory and dynamical mean-field theory approach to study both electrons and phonons in the prototypical rare-earth superhydride CeH$_9$, shedding light on the impact of electronic correlations on its critical temperature for phonon-mediated superconductivity. 
Our findings indicate that electronic correlations result in a larger electronic density at the Fermi level, a bigger superconducting gap, and softer vibrational modes associated with hydrogen atoms. 
Together, the inclusion of these correlation signatures increases the Migdal-Eliashberg 
superconducting critical temperature from 47~K to 96~K, close to the measured 95~K. 
Our results reconcile experimental observations and theoretical predictions for CeH$_9$ and herald a path towards the quantitative modeling of phonon-mediated superconductivity for interacting electron systems.
\end{abstract}
\begin{document}

\flushbottom
\maketitle
\thispagestyle{empty}
\begin{multicols}{2}
\section*{Introduction}

Pressure-induced metallic hydrogen stands as the holy grail of high-pressure physics, with a long belief in its potential room-temperature superconductivity~\cite{ashcroft1968}. 
Although the theoretical pathway is well-established, the experimental realization of metallic hydrogen is hindered by the ultra-high pressure required for metallization and the difficulty of probing samples under extreme conditions. 
A significant advancement in this field came with Ashcroft's proposal of \emph{chemical pre-compression}~\cite{ashcroft2004_i, ashcroft2004_ii}. 
This scheme widens the phase space in which a high-temperature superconductor can be found, ultimately leading to superhydrides that transition to a superconducting state at pressures lower than those predicted for elemental hydrogen~\cite{boeri2022}.

Among the various candidates, rare-earth metal superhydrides have shown great promise, especially with the discovery of LaH$_{10}$~\cite{drozdov2019,somayazulu2019evidence, Hong_2020,sun2021high}. 
It exhibits a superconducting critical temperature of around 250~K at 170~GPa, marking the highest measured superconducting critical temperature to date.
Furthermore, cerium superhydrides (CeH$_n$ with $n=9$ or $10$) have also attracted attention: although their maximum superconducting critical temperatures do not exceed that of LaH$_{10}$, they reach a relatively high superconducting critical temperature (around $100$~K) at pressures on the order of 100~GPa or less~\cite{Chen2021,cao2024, semenok2023}. 
This lower‐pressure regime not only simplifies high‐pressure synthesis and characterization but also further brings the goal of ambient‐pressure superconductivity closer to fruition.  
In general, numerous rare-earth superhydrides have been predicted and synthesized in the past decade~\cite{Peng2017, Sun2020, Guo2022, zhong2022, Zhou2020, zhou2020-2, drozdov2019, Chen2021, kong2021, shao2021, Troyan2021}, but a pivotal yet unresolved question persists.
What role do electronic correlations, induced by the $f$-electrons of the rare-earth elements, play in these high-pressure compounds?
In LaH$_{10}$, it has been observed that external pressure destabilizes the 6$s$- and 5$d$-orbitals of \ce{La}, making the $4f$-orbitals energetically more favorable~\cite{Liu2017tc}. 
As a result, the electronic states of LaH$_{10}$ near the Fermi level are composed of La $4f$- and H $1s$-orbitals~\cite{Liu2019}. 
We note that these states involving \ce{La} $4f$-orbitals are strongly localized and generally cannot be well described by density functional theory (DFT)~\cite{hohenberg1964, kohn1965}.

To gain insight into the question of the role of electronic correlation in rare-earth superhydride, we choose CeH$_9$ as a representative example and explore many-body effects using a combination method of DFT and dynamical mean field theory (DMFT), known as DFT+DMFT~\cite{anisimov2005, Kotliar2006, lechermann2006, Pourovskii2007, amadon2008, korotin2008, haule2010, haule2015, evgeny2018}. 
Previous studies have shown that the strong correlation nature of cerium $f$-electrons plays a central role in the isostructural phase transition of cerium in its elemental form~\cite{Koskenmaki1978,PhysRevLett.87.276404,PhysRevLett.96.066402,haule2007,huang2019}, in cerium intermetallic compounds~\cite{allen1986, Matar2013}, other rare-earth metals~\cite{mcmahan1998, Soderlind2014}, Kondo insulators \cite{Riseborough2000, NGCS}, as well as heavy-fermion systems~\cite{bruning2008, ohishi2209}.
We address the notable gap between the experimental results~\cite{Chen2021,Guo2022,cao2024,semenok2023} and theoretical forecasts~\cite{Peng2017, Sun2020, wang2021} regarding the critical temperature of phonon-mediated superconductivity in CeH$_9$ under high pressure.
Theoretical calculations based on DFT underestimate the superconducting critical temperature by nearly 50\%~\cite{Chen2021}, prompting speculation about the role of electronic correlations.
Our work enhances the description of phonon-mediated superconductivity by incorporating electron-correlation effects for electronic and phononic interactions.

\section*{Results}
\subsection*{Correlated electrons}

\begin{figure}[H]
\centering
\includegraphics[width=0.95\linewidth]{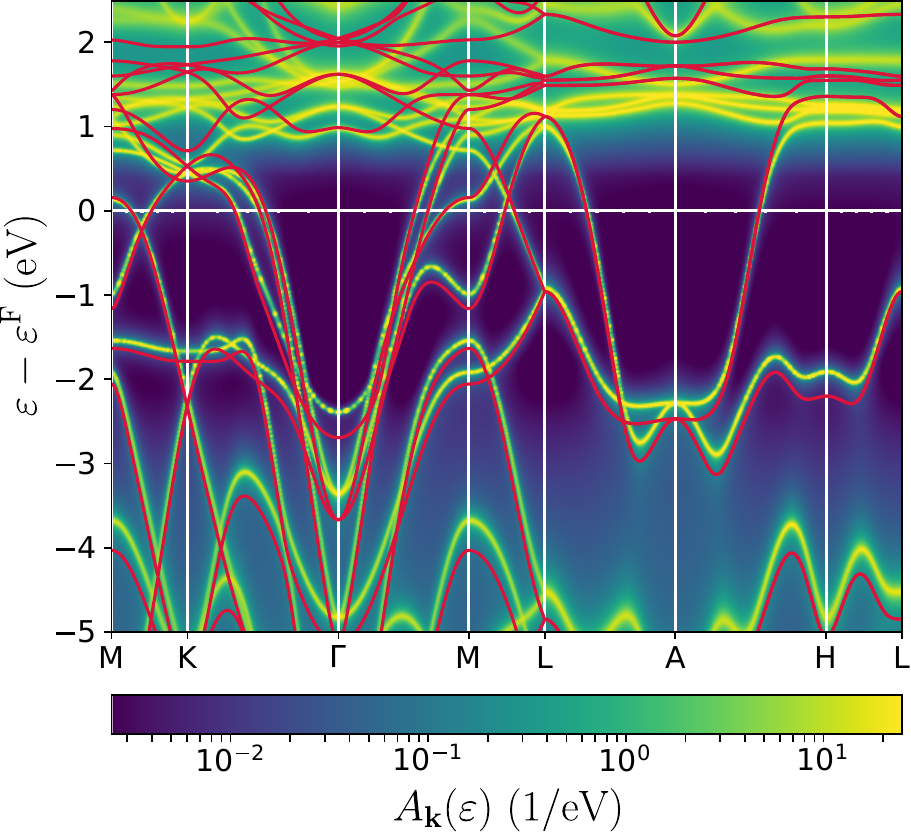}
\caption{\label{fig:ceh9-dmft-spec}
$|$ \textbf{Comparison of the electronic bands obtained from DFT and spectral function from DMFT.} Momentum-resolved spectral function $A_{\mathbf{k}}(\varepsilon)$ of CeH$_9$ along a high-symmetry path calculated with DMFT (color map) compared to the energy bands calculated by DFT (solid red lines). 
}
\end{figure}

We study the impact of electronic correlation effects in CeH$_9$ by comparing DFT results to DMFT many-body calculations with the nominal double-counting scheme~\cite{haule2007}.
To this end, we first obtain the crystal structure of CeH$_9$ from first principles by optimizing the internal atomic coordinates and the lattice constants of the crystal under 200~GPa using DFT, see Sec.~S1 in the Supplementary Information (SI).
Treating the Ce $4f$-electron as valence electrons in the pseudopotential calculations is necessary to obtain lattice parameters in agreement with measurements~\cite{jeon2020}, suggesting that the Ce-$4f$-electrons, although being more closely bound, are not atomic-like.
Turning to the electronic structure, we compare in Fig.~\ref{fig:ceh9-dmft-spec} the DFT band-structure (red lines) to the DMFT spectral function (color map).
As in LaH$_{10}$~\cite{Liu2019}, the dispersions around the Fermi level are dominated by H $1s$-orbitals that hybridize considerably with the Ce $4f$-orbitals, see Sec.~S2 of the SI for additional details.
Upon examining the intersections of the red bands and the high-intensity spectral features with the Fermi level (horizontal white line), we observe that the incorporation of electronic correlation effects results in a Fermi surface that remains largely unaltered in comparison to DFT calculations.
In addition, the multiple electron and hole pockets remain qualitatively similar throughout the Brillouin zone.
Quantitatively, however, higher-energy states are notably renormalized towards the Fermi level. 
For example, the electron pocket at the $\boldsymbol{\Gamma}$ point becomes 0.4~eV shallower as compared to the DFT result.
Notable are the energy shifts of the nondispersive Ce $4f$-orbital bands, originally 1 to 2~eV above the Fermi level, which decrease significantly in DFT + DMFT computations.

By comparing the DFT+DMFT density of state (DOS) with the DFT one in Fig.~\ref{fig:ceh9-dmft-dos}\textbf{a}, we see that the inclusion of electronic correlations shifts the prominent Ce $4f$-dominated spectral features down by at least 0.5~eV.
Since the $4f$-states are localized, their lowering causes the total spectral weight at the Fermi level to increase by 26\%, as can be seen in the inset of Fig.~\ref{fig:ceh9-dmft-dos}\textbf{a}.

We also find that DFT + U, using the fully-localized-limit double counting scheme, produces a DOS similar to DFT, see Fig.~S5 of SI, suggesting that dynamical correlations drive the observed spectral changes.
However, differences in the double-counting scheme between DFT+U and DFT+DMFT may also play a role.
Meanwhile, the orbital occupation of Ce $4f$ increases slightly from 1.42 in DFT to 1.58 with DFT+DMFT. 
This charge transfer results in effective hole-doping of the H $1s$-spectral weight. 
Wang \emph{et al.}~\cite{wang2021} noted that chemical hole doping can shift the electron pocket at $\boldsymbol{\Gamma}$ toward the Fermi level, enhancing the electron-phonon coupling. 
Remarkably, electronic correlations cause a similar effect in the stoichiometric compound.
However, despite broadenings in DFT+DMFT, the excitation spectrum remains coherent and band-like. 
This can be understood by examining the local DFT+DMFT self-energy $\Sigma_L(\varepsilon)$, from which we derive an averaged local quasiparticle weight 
$Z=\frac{1}{N_L}\sum_{L}\big[ 1 - \frac{\partial \Re \Sigma_L(\varepsilon)}{\partial \varepsilon} \big]_{\varepsilon=\varepsilon^{\rm F}} = 0.8$, where $L$ is the $4f$ manifold with $N_L=7$.
The imaginary part of the self-energy is negligible around $\pm3$~eV of the Fermi level, allowing a quasiparticle description of the low-energy spectrum.
See the Method Section and SI Sec.~S3 for more details.

\subsection*{Phonons in the presence of correlated electrons}

\begin{figure*}[ht]
\centering
\includegraphics[width=0.95\linewidth]{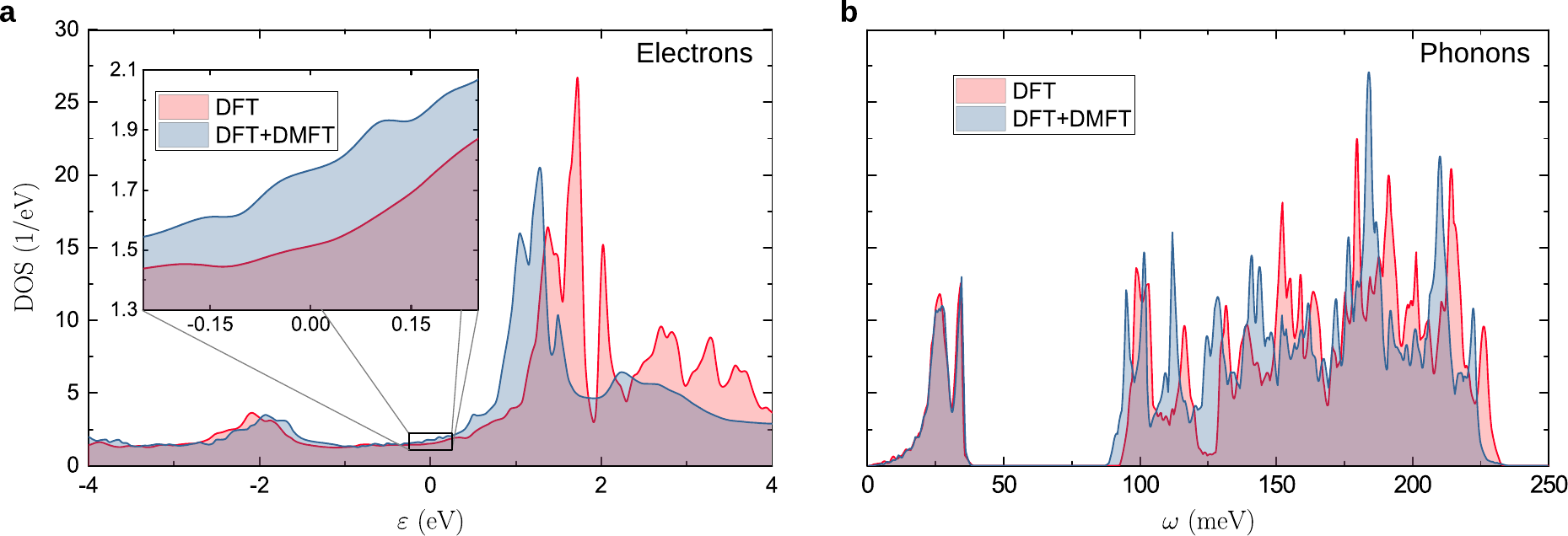}
\caption{\label{fig:ceh9-dmft-dos} 
$|$ \textbf{Comparison of density of states (DOS) per unit cell obtained from DFT and DMFT}. \textbf{a} Electronic and \textbf{b} phononic DOS of CeH$_9$ calculated with DFT and DFT+DMFT, respectively. 
The inset in \textbf{a} highlights the increase in the DOS near the Fermi level.
}
\end{figure*}

We now investigate the impact of electronic correlations on the phonon modes. 
We compute phonon dispersion via the finite difference method~\cite{parlinski1997,monserrat2018}, at the DFT level using the forces arising from small atomic displacements and at the DFT+DMFT level using forces which are instead defined from the derivatives of the free energy $F=E-TS$, where $T$ is the electronic temperature and $S$ is the impurity entropy~\cite{haule2015}. 
These calculations are enabled by the efficient methodology from Ref.~\cite{kocer2020}.
Figure~\ref{fig:ceh9-dmft-dos}\textbf{b} compares the phonon DOS of CeH$_9$ obtained within DFT and DFT + DMFT. 
Because of the significant difference in the atomic masses, the vibrations of cerium and hydrogen are largely decoupled, resulting in a clear energy separation in the phonon DOS.
The low-frequency modes ($\omega<50$ meV) arise predominantly from cerium vibrations, while the high-frequency modes ($\omega>50$ meV) emerge from dynamical distortions of the hydrogen cages. 
The highest frequency reaches $240$~meV at the DFT level (red curve), slightly lower than in LaH$_{10}$ in which the hardest mode reaches approximately 275~meV under similar pressure conditions~\cite{errea2020}. 
The effect of DMFT is to soften hydrogen vibrations.

Given that the same structure is used in both approaches, we exclude a phonon softening through volume expansion and attribute it to electronic correlations captured by DMFT. 
This can be explained by the more localized nature of DMFT electrons, which have weaker chemical bonds, and is also observed in the iron-based superconductor with inelastic neutron scattering~\cite{khanal2020}. 

\subsection*{Superconductivity}

Finally, we examine the influence of electronic correlations on the superconductivity of CeH$_9$. 
To contextualize our results, it is worth first clarifying the hierarchy of methods used to evaluate the superconducting transition temperature ($T_\mathrm{c}$): (i) the McMillan-Allen-Dynes (AD) formula ($T^\mathrm{AD}_\mathrm{c}$)~\cite{mcmillan1968, allen1975}, which provides a closed-form expression for $T_\mathrm{c}$ as follows 
\begin{figure}[H]
\centering
\includegraphics[width=0.95\linewidth]{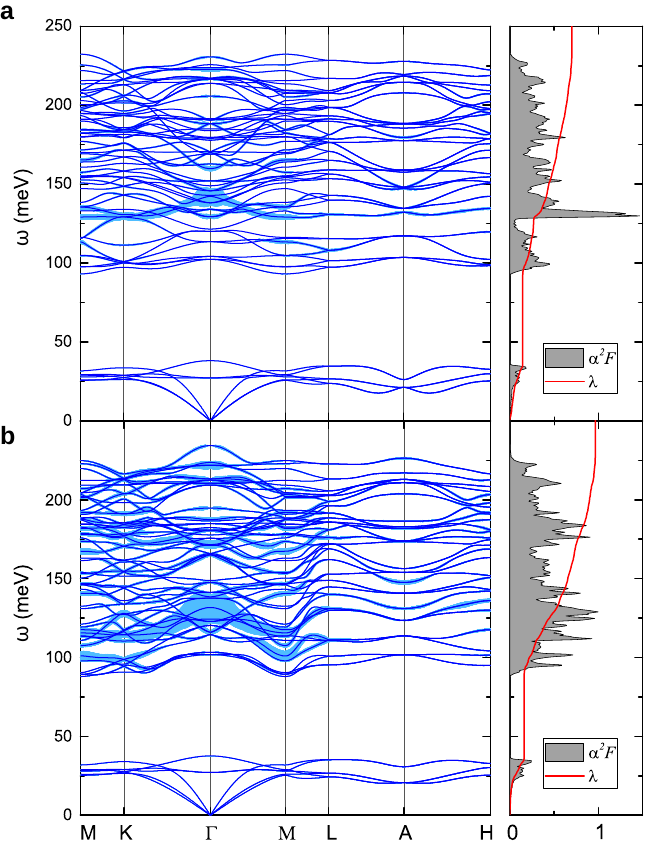}
\caption{\label{fig:ceh9-elph}  
$|$ \textbf{Phonon and electron-phonon properties obtained from DFT and DMFT}.
Phonon dispersion, Eliashberg spectral function $\alpha^2F(\omega)$, and integrated electron-phonon coupling strength $\lambda(\omega)$ as a function of phonon frequency $\omega$ as calculated with \textbf{a} DFT and \textbf{b} DFT+DMFT.
The linewidth of the phonon dispersion is proportional to $\lambda(\omega)$ projected onto the phonon modes. 
}
\end{figure}
\begin{equation}
\label{Eq.{Tc}}
T^\mathrm{AD}_\mathrm{c}=\frac{\omega_{\ln }}{1.2 k_{\mathrm{B}}} \exp \left[\frac{-1.04(1+\lambda)}{\lambda-\mu_{\mathrm{c}}(1+0.62 \lambda)}\right],
\end{equation}
\begin{figure*}[ht]
\centering
\includegraphics[width=0.95\linewidth]{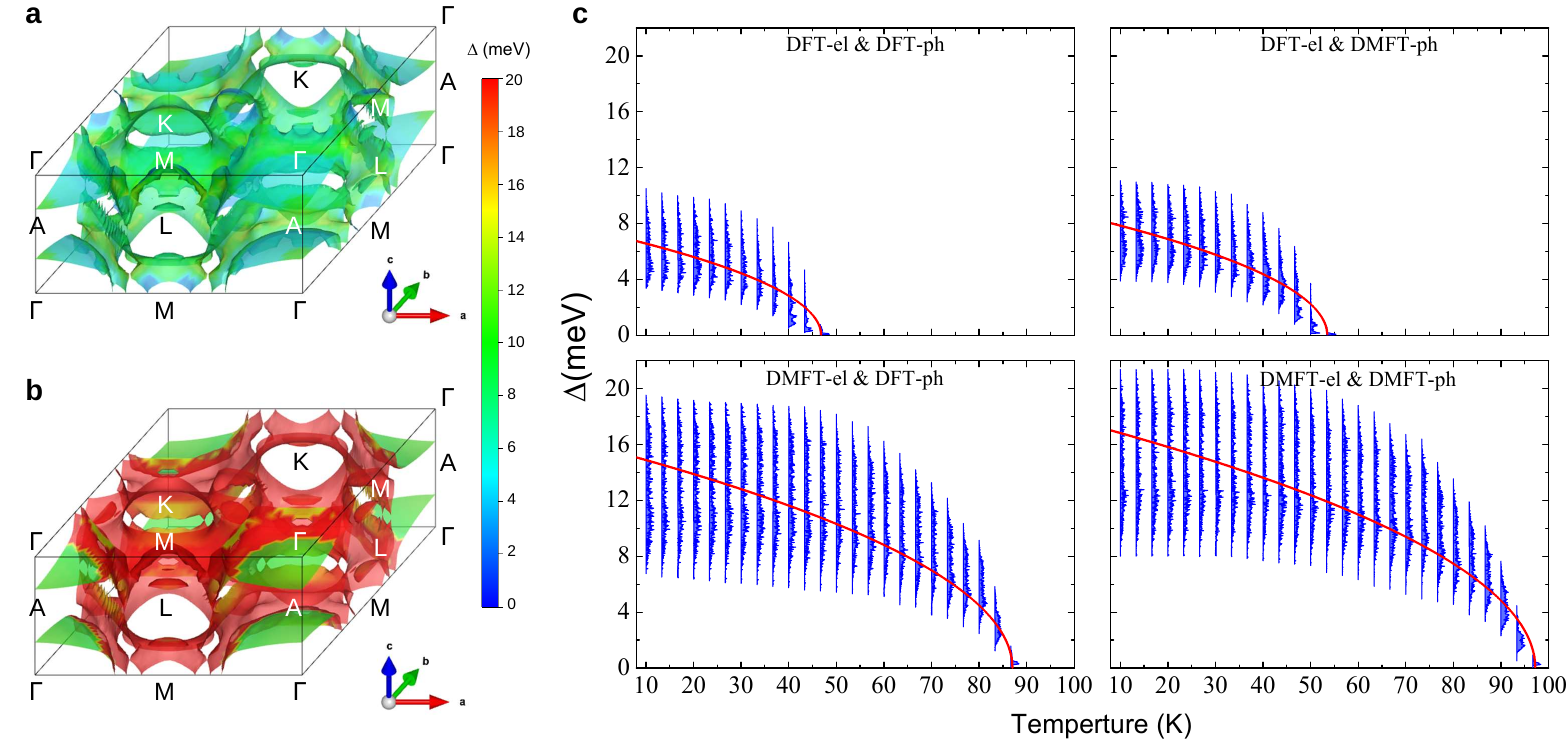}
\caption{\label{fig:Tc} 
$|$ \textbf{Anisotropic superconducting gap of CeH$_9$}.
Calculated superconducting gap $\Delta$ on the Fermi surface at 10~K computed with \textbf{a} DFT and \textbf{b} DFT+DMFT. 
\textbf{c} Anisotropic Migdal-Eliashberg superconducting gap $\Delta$ as a function of temperature. 
We consider four cases: (top left) DFT electrons with DFT phonons, (bottom left) DMFT electrons with DFT phonons, (top right) DFT electrons with DMFT phonons, and (bottom right) DMFT electrons with DMFT phonons. 
The solid red line represents the temperature dependence of the superconducting gap expected from the BCS theory~\cite{Bardeen1957} in the weak-coupling limit where $\Delta(T) = A \sqrt{1-T/T_\mathrm{c}} $, where $A$ is a fitting constant.
}
\end{figure*}
\begin{table}[H]
\centering
\resizebox{\linewidth}{!}{%
\begin{tabular}{lcccccc}
\hline \hline
     & $T_\mathrm{c}^\mathrm{aniso}$ & $T_\mathrm{c}^\mathrm{iso}$ & $T_\mathrm{c}^\mathrm{AD}$ & $P$ & $\lambda$ & $\omega_{\mathrm{ln}}$  \\
    & (K) & (K) & (K) & (GPa) &   & (meV) \\
\hline
This work  & & & & & \\
el$^{\rm DFT}$ + ph$^{\rm DFT}$   & 47 & 37 & 30 & 200 & 0.70 & 96 \\
el$^{\rm DFT}$ + ph$^{\rm DMFT}$  & 53 & 43 & 35 & 200 & 0.71 & 108 \\
el$^{\rm DMFT}$ + ph$^{\rm DFT}$  & 87 & 77 & 59 & 200 & 0.95 & 93 \\
el$^{\rm DMFT}$ + ph$^{\rm DMFT}$ & 96 & 86 & 69 & 200 & 0.97 & 106 \\
\hline
Prior works  & & & & & \\
el$^{\rm DFT}$ + ph$^{\rm DFT}$~\cite{Peng2017}  & - & 50 &  - & 100 & 0.9 & - \\
el$^{\rm DFT}$ + ph$^{\rm DFT}$~\cite{wang2021}  & - & 74  & - & 100 & 1.04 & - \\
el$^{\rm DFT}$ + ph$^{\rm DFT}$~\cite{Chen2021}  & - & 49  & 42 & 150 & 0.85 & 93 \\
el$^{\rm DFT}$ + ph$^{\rm DFT}$~\cite{Chen2021}  & - & 31  & 26 & 200 & 0.68 & 109 \\
el$^{\rm DFT}$ + ph$^{\rm DFT}$~\cite{Sun2020}  & - &  - & 74 & 250 & 0.97 & 112 \\
el$^{\rm DMFT}$ + ph$^{\rm DFT}$~\cite{plekhanov2022}   & - & - & 150 & 200 & 1.06 & 154 \\
\cline{2-4}
  experiment~\cite{Chen2021} & \multicolumn{3}{c}{95} &  110 & - & - \\
  experiment~\cite{Chen2021} & \multicolumn{3}{c}{95} &  150 & - & - \\
  experiment~\cite{cao2024} & \multicolumn{3}{c}{110} &  110 & - & - \\
  experiment~\cite{semenok2023} & \multicolumn{3}{c}{90} &  116 & - & - \\
\hline \hline
\end{tabular}
}
\newline
\newline
\caption{\label{tab:ceh9-tc}
\textbf{Critical temperatures and superconducting parameters for CeH$_9$}.
We report the logarithmic averaged phonon frequency ($\omega_\mathrm{ln}$) and electron-phonon coupling strength ($\lambda$) entering the Allen-Dyne $T^\mathrm{AD}_\mathrm{c}$, isotropic Migdal-Eliashberg $T^\mathrm{iso}_\mathrm{c}$, and anisotropic Migdal-Eliashberg $T^\mathrm{aniso}_\mathrm{c}$. 
Our calculations describe the electron-phonon pairing on four levels of sophistication and are compared to previous theoretical works at the DFT level and experimental measurements.
All theoretical values of $T_\mathrm{c}$ reported in the table are calculated with the Morel-Anderson pseudopotential $\mu_\mathrm{c}=0.13$ except Ref.~\cite{Chen2021} which uses $\mu_\mathrm{c}=0.15$.
}
\end{table}
\noindent where $\omega_{\ln}$ is the logarithmic averaged phonon frequency, $k_\mathrm{B}$ denotes the Boltzmann constant, $\lambda$ is the electron-phonon (el-ph) coupling strength, and $\mu_\mathrm{c}$ is the Morel-Anderson pseudopotential accounting for Coulomb repulsion of the system~\cite{Morel1962}; (ii) the isotropic ($T^\mathrm{iso}_\mathrm{c}$) and (iii) anisotropic Eliashberg equations ($T^\mathrm{aniso}_\mathrm{c}$)~\cite{allen1982,margine2013}, in which $T_\mathrm{c}$ is determined by identifying the highest temperature at which a nonzero superconducting gap $\Delta$ appears. 
These three methods provide a progressively more refined description of phonon-mediated superconductivity with increasing accuracy and computational cost.

We note that Plekhanov \emph{et al.}~\cite{plekhanov2022} have used the AD formula to obtain a correlation-induced enhancement in $T_\mathrm{c}$, with the many-body correction introduced directly at the level of the Eliashberg spectral function $\alpha^2F(\omega)$. 
They obtain a $T^\mathrm{AD}_\mathrm{c} = 150$~K at 200~GPa, which overestimates the experimental value of 95~K~\cite{Chen2021}.  
This overestimation may be attributed to the use of the atomic Hubbard-I~\cite{hubbard1963} solver for the DMFT impurity model and the use of DFT phonons. 
Our approach overcomes these limitations through the use of the continuous-time quantum Monte Carlo solver~\cite{haule2007,Gull2011} which is unbiased towards electron localization, and by ensuring an equal-footing treatment of DMFT electrons and phonons.

Interestingly, with both electrons and phonons described at the DFT level, we find that the el-ph coupling strength for CeH$_9$ is $\lambda=2 \int_{0}^{\infty} \mathrm{d} \omega \, \alpha^{2} F(\omega)/{\omega} = 0.70$. 
We then compute the superconducting critical temperatures from the AD formula finding $T^\mathrm{AD}_\mathrm{c}\simeq 30$~K, and by solving the isotropic and anisotropic Eliashberg equations, leading to $T^\mathrm{iso}_\mathrm{c}\simeq37$~K and $T^\mathrm{aniso}_\mathrm{c}\simeq47$~K, respectively. 
As anticipated, the McMillan–Allen–Dynes estimate is the lowest, since that formula is formally valid only for low-$T_\mathrm{c}$ superconductors. 
Our values for $\lambda, $ $T_\mathrm{c}^\mathrm{AD}$ and $T_\mathrm{c}^\mathrm{iso}$ agree well with a previous study~\cite{Chen2021}. 
The Eliashberg treatment improves the prediction. 
Importantly, the anisotropic Eliashberg solution demonstrates that anisotropy significantly increases the predicted $T_\mathrm{c}$, bringing it closer to experimental observations.

Figure~\ref{fig:ceh9-elph}\textbf{a} shows the integrated el-ph coupling strength $\lambda$($\omega$)=$2\int_{0}^{\omega} \mathrm{d} \omega \, \alpha^{2} F(\omega)/{\omega}$, el-ph coupling strength projected onto the phonon dispersion, $\lambda_{\mathbf{q}\nu}$, and  the Eliashberg function $\alpha^{2} F(\omega)$. 
We find that $\lambda_{\mathbf{q}\nu}$ is distributed in a nonuniform manner among the phonon modes and the phonon Brillouin zone, with approximately 80\% of the total el-ph coupling strength arising from the optical phonon modes of hydrogen and the remaining 20\% from the phonon modes of cerium. 
In particular, as highlighted by the linewidth in Fig.~\ref{fig:ceh9-elph}\textbf{a}, phonon modes with frequency around 125~meV are highly coupled to electrons, leading to a significant peak in the Eliashberg function $\alpha^2F(\omega)$. 
This frequency range is also where we observed a significant softening of the hydrogen mode by electronic correlations.

To go further, we substitute the electron Hamiltonian and phonon dynamical matrices obtained at the DFT level with their counterparts at the DMFT level.
These two substitutions affect the electronic dispersion, the phonon dispersion, and the el-ph coupling matrix elements.
We update the latter via the DMFT phonon eigendisplacement and frequencies but we neglect the effect of DMFT on the perturbed electronic potential. 
Current implementations of the Migdal-Eliashberg theory of superconductivity are based on the quasiparticle approximation. 
Therefore, we construct from our DFT+DMFT calculations an effective quasiparticle band structure for CeH$_9$. 
The construction of a renormalized band-structure is possible as the real-part of the self-energy can, to good approximation, be linearlized around the Fermi level where its imaginary part is small.
Then, an effective band structure can be found 
by upfolding the local DMFT self-energy into the Kohn-Sham basis $\Sigma_{nn'\mathbf{k}}(\varepsilon) = \sum_L P_{nn'L\mathbf{k}} \Sigma_L(\varepsilon)$ where $P_{nn'L\mathbf{k}}$ is the inverse of the projector that defines the local $4f$ basis~\cite{haule2010}.
In the DFT+DMFT, the charge density is obtained self-consistently, leading to an updated diagonal Kohn-Sham Hamiltonian
$H^{\rm DFT+DMFT}_{nn\mathbf{k}}$.
We obtain the renormalized quasiparticle energies $E_{n\mathbf{k}}$ by solving the following linearized equation~\cite{Tomczak2007}: 
\begin{equation}
\operatorname{det} \left( E_{n\mathbf{k}} \delta_{nn'} - \sum_m Z_{nm\mathbf{k}} \left [H_{mn'\mathbf{k}}^{\mathrm{\tiny DFT+DMFT}} + \Re  \Sigma_{mn'\mathbf{k}}(\varepsilon^{\rm F}) \right ]  \right)  = 0,
\end{equation}
where 
\begin{equation}
Z_{nn'\mathbf{k}} = \bigg[ 1 - \frac{\partial \Re \Sigma_{nn'\mathbf{k}}(\varepsilon)}{\partial \varepsilon}\Big|_{\varepsilon=\varepsilon^{\rm F}}\bigg]^{-1}    
\end{equation}
is the quasiparticle weight which approaches unity in a noninteracting system and zero in the limit of extremely strong correlations.

We validate the quality of the effective band structure obtained using the above quasiparticle approximation in Sec.~S4 of the SI.
Using this effective band structure and the DMFT phonon spectrum obtained previously, we calculate the el-ph coupling strength and the Eliashberg function at the DMFT level, as shown in Fig.~\ref{fig:ceh9-elph}\textbf{b}. 
The total coupling strength in this case is found to increase by nearly 39\%, reaching $\lambda\simeq0.97$. 
This increase supports a substantial elevation in the critical temperature, which increases from $T_\mathrm{c}^\mathrm{AD} =30$~K at the DFT level to $T_\mathrm{c}^\mathrm{AD}=69$ K at the DMFT level. 
We traced the increase in $T_\mathrm{c}$ to changes in specific regions in the Brillouin zone.

Figures~\ref{fig:Tc}\textbf{a}-\textbf{b} compare the superconducting gap $\Delta$ on the Fermi surface of CeH$_9$ at 10~K. 
We find that in the case of DFT the superconducting gap is quite uniform across the Fermi surface, with an average magnitude of approximately 6.3~meV that varies by $\pm 2.5$ meV through the Brillouin zone. 
In contrast, at the DMFT level, the superconducting gap around the corner of the Brillouin zone (\ie the $K$ point) exhibits a remarkable increase, up to 20~meV, which explains the elevated superconducting critical temperature, from $T_\mathrm{c}^\mathrm{aniso} =47$~K at the DFT level to $T_\mathrm{c}^\mathrm{aniso}=96$~K at the DMFT level.

To gain further insights into the role of electronic correlation, we consider two `intermediate scenarios' of el-ph coupling: DMFT electrons are paired with DFT phonons, and DFT electrons are paired with DMFT phonons. 
Figure~\ref{fig:Tc}\textbf{c} shows the energy distribution of the superconducting gap $\Delta$ as a function of temperature for our previous results at the DFT and DMFT levels, as well as for the two intermediate scenarios.
The superconducting gap is seen to close gradually with increasing temperature, vanishing at various critical values for each case.
Table~\ref{tab:ceh9-tc} summarizes the critical temperatures obtained, together with the corresponding superconducting parameters, offering a hierarchical comparison also with existing DFT studies and experimental measurements. 

Our results align closely with previous DFT studies but overestimate experimental values at comparable pressures by 50\%. 
In contrast, our value at the DMFT level, $T_\mathrm{c}^\mathrm{aniso} = 96$~K at 200~GPa, is in excellent agreement with the experimental value, $T_\mathrm{c} = 95$~K at the closest pressure of 150~GPa. 
Dissecting the influence of electronic correlations on the different ingredients entering the calculation of $T_\mathrm{c}$, we find that the DMFT renormalization of the electronic structure is the main driving force behind the increased critical temperature. 
Indeed, keeping the phonons at the DFT-level, the effect of DMFT on electrons almost doubles the $T_\mathrm{c}$.
Indeed, as shown in Fig~\ref{fig:ceh9-dmft-dos}\textbf{a}, the increased DOS at the Fermi level, $D(\varepsilon^\mathrm{F})$, facilitates the formation of more Cooper pairs, leading to a higher superconducting critical temperature for which $T_\mathrm{c} \propto \exp{(-\frac{1}{V D(\varepsilon^\mathrm{F})})}$ where $V$ is the pairing interaction~\cite{Bardeen1957}. 
Finally, we also discover that exclusively renormalizing the electronic structure with DMFT still leads to a substantial underestimation of the superconducting critical temperature. 
It is necessary to also include the phonon-softening effect due to electronic correlations to quantitatively reconcile the theoretical findings with experimental data.

\section*{Discussion}

In summary, we have used a combination of DFT and DFMT to study electrons and phonons of CeH$_9$ under high pressure, demonstrating that strong electronic correlations arising from cerium $4f$-states play an important role in this system. 
Many-body effects increase electron masses, produce a larger spectrum at the Fermi level, and soften
hydrogen vibrational modes.
Combining DFT+DMFT with the Migdal-Eliashberg theory of superconductivity, we were able to investigate the role of electronic correlations for the critical temperatures of phonon-mediated superconductivity in CeH$_9$. 
Our findings indicate that the dominant correction to the critical temperature comes from the many-body renormalization of the electronic structure. 
However, we find that the renormalization of phonons is significant and cannot be neglected. 
The many-body-driven increase in the critical superconducting temperature is 50~K.
Future extensions of this work should address the impact of many-body correlations on the perturbed potential, given that exchange and correlation effects beyond semi-local DFT can either increase~\cite{PhysRevB.105.014517,PhysRevX.3.021011,mandal2014,zou2025,Poliukhin2025} or decrease~\cite{Yam2022,Coulter2025} the values of the el-ph coupling matrix elements.

Overall, the renormalized Migdal-Eliashberg approach developed here is not specific to CeH$_9$ but can be applied to other compounds where el-ph interactions coexist with electronic correlations. 
Promising targets include transition-metal hydrides with partially filled $d$-shells, such as X$_2$MH$_6$ (where X are alkali, alkaline-earth, or post-transition metals and M are 3\textit{d}-, 4\textit{d}-, and 5\textit{d}-transition metals) which have recently attracted attention for their potential high-temperature superconductivity at ambient pressure~\cite{Zheng2024}. 
In addition, while our approach focuses on phonon-mediated superconductivity, it could serve as a reference point to assess the relative importance of el-ph coupling effects in materials where other mechanisms such as spin-fluctuations~\cite{Kawamura2020,Worm2024} dominate. 
Future developments incorporating electronic self-energy effects beyond the Migdal-Eliashberg theory may help clarify how phonons influence unconventional superconductivity in these systems.

\section*{Methods}
\subsection*{Electronic structure calculations}
We have performed DFT + DMFT calculations at inverse temperature $\beta=20$~eV$^{-1}$ (\ie$T \simeq 580$~K) with the embedded DMFT functional package \textsc{eDMFT}~\cite{haule2015} integrated with the all-electron DFT code \textsc{WIEN2k}~\cite{blaha2020} for both electrons and phonons in CeH$_9$. 
A single calculation at a lower temperature with $\beta=100$~eV$^{-1}$ (\ie$T \simeq 116$~K) has also been performed to examine possible effects of temperature onto the electronic self-energy. The resulting self-energy was found to fall on top of that obtained at $\beta = 20~\mathrm{eV}^{-1}$ (see Fig.~S6\textbf{a}).
The DMFT impurity problem is solved using a continuous-time quantum Monte Carlo algorithm~\cite{haule2007,Gull2011}. 
We use the nominal double counting scheme~\cite{haule2007}, with an occupation fixed to 2, as the DFT-calculated valency of Ce under pressure is found to be 1.42 (greater than the nominal 1 electron) indicating an increased $f$-electron occupancy compared to the idealized trivalent state. 
The impact of the choice of the nominal occupation values on the self-energy has been examined in Sec.~S5 of the SI, where the quasiparticle renormalization factor is found to be independent when the nominal occupation varies from 1 to 2.
The corresponding DMFT occupation is increasing from 1.11 to 1.58 for the same range. The standard nominal occupation for (unpressurized) Ce-compounds, $n_f^0=1$, corresponds to a Ce$^{3+}$ configuration for pressurized CeH$_9$, which we find implausible given the DFT valency for CeH$_9$.
We also note that the magnetic susceptibility of CeH$_9$ exhibits no long-term memory, see SI Sec.~S6, indicating the absence of fluctuating local moments in CeH$_9$.
Details of Haule \emph{et al.}'s implementation can be found in Ref.~\cite{haule2015}.
We analytically continue the Matsubara self-energy to real frequencies using the maximum entropy method~\cite{kaufmann2023}.
The quasiparticle weight was extracted from both Matsubara and real-frequency data and was found to be consistent.
We have cross-checked the \textsc{Wien2k} all-electron results with the pseudopotential-based DFT packages \textsc{CASTEP}~\cite{segall2002, clark2005} and \textsc{Quantum ESPRESSO}~\cite{giannozzi2009, giannozzi2017} that were used previously in Ref.~\cite{plekhanov2022}, using the same ultrasoft pseudopotentials that \textsc{CASTEP} generates on the fly for both codes. 
The valence electron configurations of $5s^25p^64f^15d^16s^2$ (with $4f$-electrons explicitly included) and $1s$ are adopted for the Ce and H atoms, respectively. 
Figure~S11 in SI Sec.~S7 shows close agreement between the three codes. 
Convergence tests at the DFT level yielded an energy cut-off of 800~eV for \textsc{CASTEP} and for \textsc{Quantum ESPRESSO}. 
For \textsc{Wien2k}, convergence is achieved when the product of the smallest atomic‐sphere radius and the maximum reciprocal‐space vector in the plane‐wave expansion reaches 7.
A $\boldsymbol{\Gamma}$-centered $\mathbf{k}$-point grid of size $12 \times 12 \times 6$ is used for all three packages. 
The exchange-correlation functional is approximated using PBE~\cite{perdew1996generalized}.
As was done in previous works~\cite{Peng2017, Chen2021, wang2021, Sun2020, plekhanov2022}, we omit spin-orbit coupling for technical reasons.
In the DFT + DMFT calculations, we use an on-site Coulomb interaction (Hubbard $U$) of $U=6.0$~eV and a Hund's coupling $J=0.7$~eV for the cerium $f$ orbitals. 
This choice is comparable to previous values that were successfully used for elemental cerium~\cite{PhysRevLett.87.276404,PhysRevLett.96.066402,haule2007,huang2019}, CeH$_9$~\cite{plekhanov2022}, and other Ce-compounds~\cite{NGCS}.
Note that we neglect any potential impact of pressure on Hubbard $U$~\cite{mcmahan1998,jmt_wannier}.
We also omit any additional renormalizations of the DMFT electronic structure that arise from the electron-phonon coupling~\cite{abramovitch2023,abramovitch2025}.

\subsection*{Lattice dynamics calculations}
For the lattice dynamics calculations, we use finite difference with an in-house code which offers a versatile interface with DFT and DFT+DMFT~\cite{kocer2020}, to calculate the phonon spectrum of CeH$_9$. 
Considering that the computational cost of DFT+DMFT exhibits a particularly unfavorable scaling with the system size, we employ the non-diagonal supercell technique~\cite{lloyd2015} to sample a $2 \times 2 \times 2$ $\mathbf{q}$-point grid in the Brillouin zone and then interpolate the result to obtain dynamical matrices on a $\mathbf{q}$-point grid of $6 \times 6 \times 3$.
A convergence validation of the phonon dispersions with the size of the $\mathbf{q}$-point grid at the DFT level is provided in Sec.~S8 of the SI.
In addition, we find that the fixed self-energy approximation~\cite{kocer2020} is not applicable to CeH$_9$, see Sec.~S9 of the SI, and thus fully self-consistent DFT+DMFT calculations are performed for each displaced configuration.
Specifically, we use four different supercells constructed with the following supercell matrices:  $\big(\begin{smallmatrix}
  1 & 0 & 0\\
  0 & 1 & 0\\
  0 & 0 & 1\\
\end{smallmatrix}\big)$, $\big(\begin{smallmatrix}
  1 & 0 & 0\\
  0 & 1 & 0\\
  0 & 0 & 2\\
\end{smallmatrix}\big)$, $\big(\begin{smallmatrix}
  1 & 0 & 0\\
  0 & 2 & 0\\
  0 & 0 & 1\\
\end{smallmatrix}\big)$, and $\big(\begin{smallmatrix}
  1 & 0 & 0\\
  0 & 1 & -1\\
  1 & 2 & 0\\
\end{smallmatrix}\big)$, the first one provides access to $\boldsymbol{\Gamma}-\mathbf{q}$-point, then the second one provides access to the $(0,0,1/2)$ $\mathbf{q}$-point, the third one to the $(0,1/2,0)$ $\mathbf{q}$-point, and the forth one to the $(1/2,1/2,1/2)$ $\mathbf{q}$-point. 
All other $\mathbf{q}$-points on the $2 \times 2 \times 2 $ grid are related to these four by symmetry and therefore explicit calculations are not necessary. 

\subsection*{Superconductivity calculations}
To evaluate the superconducting critical temperature $T_\mathrm{c}$, we used \textsc{EPW}~\cite{ponce2016, lee2023} to interpolate the el-ph coupling matrix calculated with the density functional perturbation theory as implemented in \textsc{Quantum ESPRESSO} from a $12 \times 12 \times 6$ $\mathbf{k}$-point grid and a $6 \times 6 \times 3$ $\mathbf{q}$-point grid to a $48 \times 48 \times 24$ $\mathbf{k}$-point grid and a $24 \times 24 \times 12$  $\mathbf{q}$-point grid. 
In particular, for $T_\mathrm{c}$ calculations at the DFT+DMFT level, the electronic and phononic degrees of freedom obtained from standard DFT are replaced with the counterparts obtained from DFT+DMFT.
We used a standard value for the Morel-Anderson pseudopotential $\mu_\mathrm{c} = 0.13$ which accounts for Coulomb repulsion of the system~\cite{Morel1962} in all our calculations. 

\section*{Data availability}
The data and codes that support the findings of this study are available in the Supplementary Information and on the Materials Cloud Archive at \href{https://doi.org/10.24435/materialscloud:hw-nk}{https://doi.org/10.24435/materialscloud:hw-nk}.

\section*{Code availability}
The \textsc{Wien2k} v21.1 code used in this study is a commercial software, available from \href{http://susi.theochem.tuwien.ac.at/}{http://susi.theochem.tuwien.ac.at}.
The \textsc{Quantum ESPRESSO} v7.0 code used in this research is open source: \href{https://www.quantum-espresso.org}{https://www.quantum-espresso.org}. 
The \textsc{CASTEP} 23.1 code used in this study is freely available at \href{https://www.castep.org}{https://www.castep.org} for academic research. 
The \textsc{EPW} v5.4.1 code used in this research is open source: \href{https://epw-code.org}{https://epw-code.org}. 
The \textsc{ana\_cont} v1.1.2 code used in this study is open source: \href{https://github.com/josefkaufmann/ana\_cont}{https://github.com/josefkaufmann/ana\_cont}.
Haule \emph{et al.} provide their DMFT package (Oct-2024 release) at \href{https://github.com/ru-ccmt/eDMFT}{https://github.com/ru-ccmt/eDMFT}.

\section*{Authors contributions}

S.C. and Y.W. contributed equally to this work.
S.C. performed the DMFT spectral function and superconducting calculations, while Y.W. performed the DMFT phonon calculations. 
S.P. and J.M.T provided technical help with the \textsc{Wien2k}, DMFT and \textsc{EPW} software. 
S.P., B.M. and J.M.T supervised the work.
All authors reviewed and approved the final version of the manuscript.

\section*{Competing interests}

The authors declare no competing interests.

\bibliography{references}

\section*{Acknowledgements}
The authors would like to thank C. Weber and E. Plekhanov for useful discussions. 
S.C. and B.M. acknowledge support from EPSRC [EP/V062654/1]. 
S.C. also acknowledges financial support from the Cambridge Trust and the Winton Program for the Physics of Sustainability. 
Y.W. acknowledges funding from the China Scholarship Council. 
B.M. also acknowledges support from a UKRI Future Leaders Fellowship [MR/V023926/1] and from the Gianna Angelopoulos Programme for Science, Technology, and Innovation.
S.P. is a Research Associate of the Fonds de la Recherche Scientifique - FNRS.
This publication was supported by the Walloon Region in the strategic ax FRFS-WEL-T.
Computational resources were provided through our membership in the UK's HEC Materials Chemistry Consortium, funded by EPSRC (EP/R029431 and EP/X035859). 
This work used ARCHER2, the UK National Supercomputing Service (http://www.archer2.ac.uk), as well as resources from the UK Materials and Molecular Modeling Hub (MMM Hub), which is partially supported by EPSRC (EP/T022213 and EP/W032260).
Computational resources were also provided by the EuroHPC JU award granting access to MareNostrum5 at the Barcelona Supercomputing Center (BSC), Spain (Project ID: EHPC-EXT-2023E02-050), by the Consortium des Équipements de Calcul Intensif (CÉCI), funded by the FRS-FNRS under Grant No. 2.5020.11, by the Tier-1 supercomputer of the Walloon Region (Lucia) with infrastructure funded by the Walloon Region under the grant agreement n°1910247.

\end{multicols}
\end{document}


\title{Supplemental Information for\\ Impact of strong correlation on the superconductivity of the high-pressure rare-earth \ce{CeH9}}

\author{Siyu Chen}
\affiliation{TCM Group, Cavendish Laboratory, University of Cambridge,
J. J. Thomson Avenue, Cambridge CB3 0HE, United Kingdom}
\affiliation{Department of Materials Science and Metallurgy, University of Cambridge, 27 Charles Babbage Road, Cambridge CB3 0FS, United Kingdom}
\affiliation{European Theoretical Spectroscopy Facility, Institute of Condensed Matter and Nanosciences, Université catholique de Louvain, Chemin des Étoiles 8, B-1348 Louvain-la-Neuve, Belgium}
\author{Yao Wei}
\affiliation{Department of Physics, King’s College London, Strand, London WC2R 2LS, United Kingdom}
\author{Bartomeu Monserrat}
\affiliation{TCM Group, Cavendish Laboratory, University of Cambridge,
J. J. Thomson Avenue, Cambridge CB3 0HE, United Kingdom}
\affiliation{Department of Materials Science and Metallurgy, University of Cambridge, 27 Charles Babbage Road, Cambridge CB3 0FS, United Kingdom}
\author{Jan Tomczak}
\affiliation{Department of Physics, King’s College London, Strand, London WC2R 2LS, United Kingdom}
\affiliation{Institute of Solid State Physics, TU Wien, 1040 Vienna, Austria}
\author{Samuel Ponc\'e}
\affiliation{European Theoretical Spectroscopy Facility, Institute of Condensed Matter and Nanosciences, Université catholique de Louvain, Chemin des Étoiles 8, B-1348 Louvain-la-Neuve, Belgium}
\affiliation{WEL Research Institute, avenue Pasteur 6, 1300 Wavre, Belgium}

\maketitle

\section{Section S1: Crystal structure of high-pressure $\text{CeH}_9$}
\begin{figure}[H]
\centering
\includegraphics[width=0.75\linewidth]{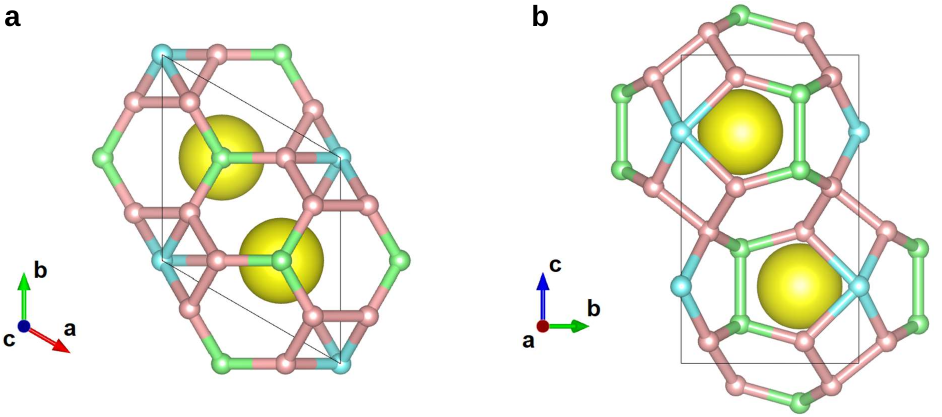}
\caption{(a) Top and (b) side views of the crystal structure of \ce{CeH9}, where the cerium atom is inside the cage formed by hydrogen atoms. The hydrogen atoms with different chemical environments are distinguished in red, green and blue.}
\label{fig:ceh9-structure}
\end{figure}

The high-pressure phase of \ce{CeH9} was first theoretically predicted~\citep{Peng2017}. 
Subsequent experimental studies have confirmed its thermodynamic and stoichiometric stability under external pressure exceeding 80\,GPa~\citep{salke2019, li2019_ceh9}. 
The structure is shown in Fig.\,\ref{fig:ceh9-structure} crystallises in the space group P$\mathrm{6_3/mmc}$, where the cerium atoms form a hexagonal close-packed lattice. 
Each cerium atom is surrounded by 29 hydrogen atoms, and these hydrogen atoms can be considered as an octadecahedronal cage composed of six tetragonal rings, six pentagonal rings and six hexagonal rings. 
The rings share edges between contiguous octadecahedral cages, giving rise to an extensive 3D hydrogen network structure. 
It is this special clathrate structure that allows high hydrogen content without the formation of H$_{2}$ molecules. 

There are three chemically non-equivalent hydrogen sites in \ce{CeH9} (highlighted in red, green and blue in Fig.\,\ref{fig:ceh9-structure}), thereby the hydrogen atoms have four nearest-neighbour distances.  
Table \,\ref{tab:ceh9_structure} reports for the structural parameters of the hydrogen cage under the pressure of 200 GPa.
Among them, we find that the `red' and `green' hydrogen atoms exhibit the shortest distance of 1.122\,Å.
This distance is comparable to that in metallic hydrogen at 500\,GPa~\citep{azadi2014}. 
In this sense, cerium provides a chemical precompression pressure of nearly 300\,GPa. 
\vspace{1cm}

\begin{table}[H]
\centering
\resizebox{\textwidth}{!}{%
\begin{tabular}{@{}ccccccccc@{}}
\hline
\hline
Source & Pressure &  Space Group &  Lattice Parameters  &  Atomic Sites & $x$ & $y$ & $z$ &  Nearest-neighbor distances\\
\hline
This work & 200~GPa & P$6_3/mmc$ & \begin{tabular}[c]{@{}c@{}}\quad $a = 3.49$ \si{\angstrom} \\\quad $c = 5.23$ \si{\angstrom} \end{tabular} &
\begin{tabular}[c]{@{}l@{}}\quad Ce\\ \quad H$_{\mathrm{r}}$\\ \quad H$_{\mathrm{g}}$\\ \quad H$_{\mathrm{b}}$\end{tabular} & 
\begin{tabular}[c]{@{}l@{}}2/3\\ 0.155\\ 1/3\\ 0\end{tabular} & 
\begin{tabular}[c]{@{}l@{}}1/3\\ 0.309\\ 2/3\\ 0\end{tabular} & 
\begin{tabular}[c]{@{}l@{}}1/4\\ 0.071\\ 0.129\\ 3/4\end{tabular} & 
\begin{tabular}[c]{@{}c@{}} \quad \quad H$_{\mathrm{r}}$-H$_{\mathrm{r}}$ = 1.200 \si{\angstrom} \\ \quad \quad H$_{\mathrm{r}}$-H$_{\mathrm{g}}$ = 1.122 \si{\angstrom}\\ \quad \quad H$_{\mathrm{r}}$-H$_{\mathrm{b}}$ = 1.320 \si{\angstrom}\\ \quad \quad H$_{\mathrm{g}}$-H$_{\mathrm{g}}$ = 1.264 \si{\angstrom}\end{tabular} \\
Experimental~\cite{li2019_ceh9} & 160~GPa &  P$6_3/mmc$ & \begin{tabular}[c]{@{}c@{}}\quad $a = 3.58$ \si{\angstrom} \\\quad $c = 5.41$ \si{\angstrom} \end{tabular} & \begin{tabular}[c]{@{}l@{}}\quad Ce\\ \quad H$_{\mathrm{r}}$\\ \quad H$_{\mathrm{g}}$\\ \quad H$_{\mathrm{b}}$\end{tabular} & 
\begin{tabular}[c]{@{}l@{}}2/3\\ 0.156\\ 1/3\\ 0\end{tabular} & 
\begin{tabular}[c]{@{}l@{}}1/3\\ 0.312\\ 2/3\\ 0\end{tabular} & 
\begin{tabular}[c]{@{}l@{}}1/4\\ 0.062\\ 0.149\\ 1/4\end{tabular} & 
\begin{tabular}[c]{@{}c@{}} \quad \quad H$_{\mathrm{r}}$-H$_{\mathrm{g}}$ = 1.092 \si{\angstrom}\\ \end{tabular} \\
\hline
\hline
\end{tabular}
}
\caption{Lattice parameters and atomic positions for the \ce{CeH9} crystallizing in the P$\mathrm{6_3/mmc}$ space group, including lattice parameters, atomic sites, fractional coordinates ($x$, $y$, $z$), and nearest-neighbor distances.}
\label{tab:ceh9_structure}
\end{table}

\vspace{1cm}
\section{Section S2: Density of states and spectral functions }
\begin{figure}[H]
\centering
\includegraphics[width=0.46\linewidth]{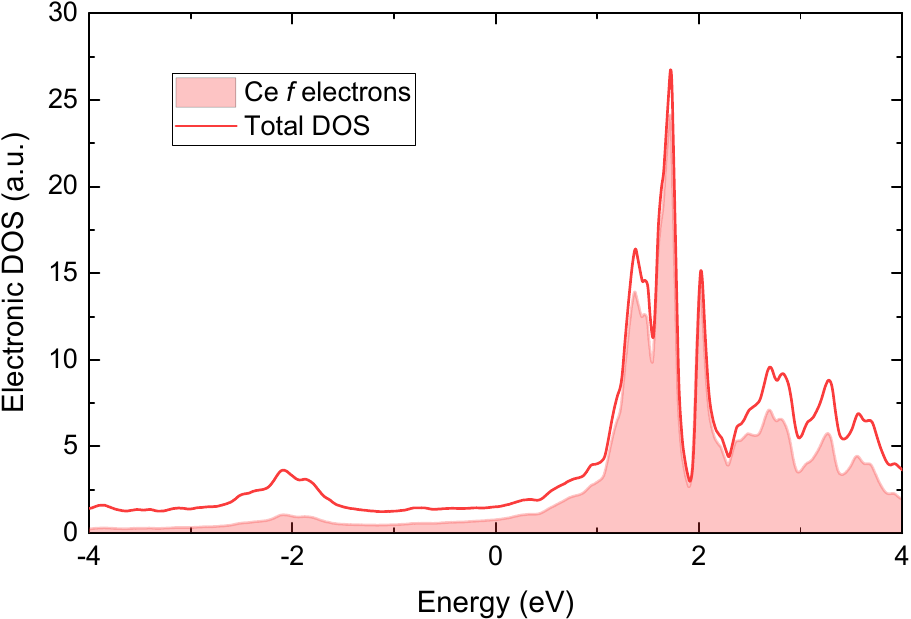}
\caption{Density of states (DOS) of \ce{CeH9} calculated through DFT, where the contribution of cerium $f$-electrons is highlighted.}
\label{fig:ceh9-dos}  
\end{figure}

\begin{figure}[H]
\centering
\includegraphics[width=0.46\linewidth]{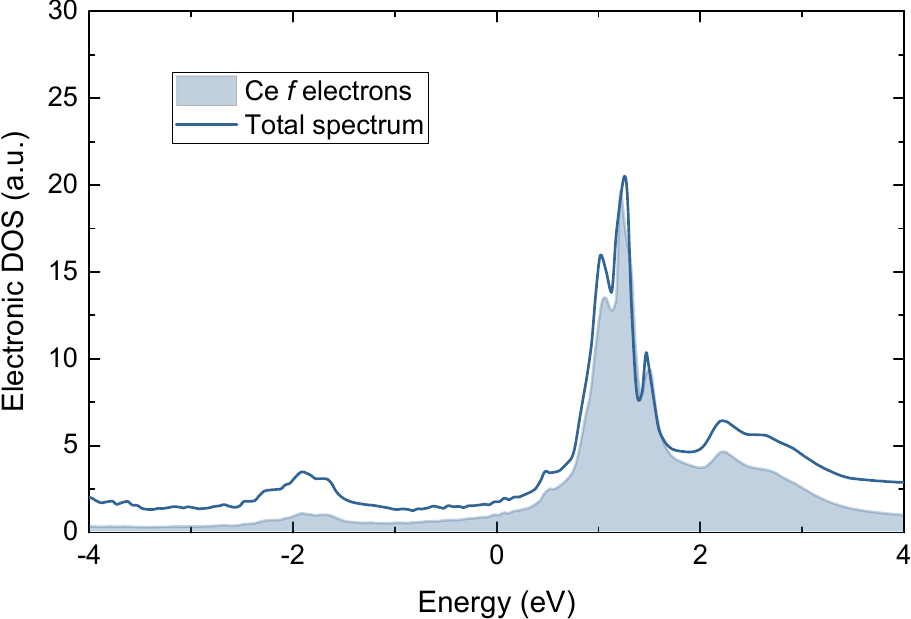}
\caption{Density of states (DOS) of \ce{CeH9} calculated through DFT+DMFT, where the contribution of cerium $f$-electrons is highlighted.}
\label{fig:ceh9-dos}  
\end{figure}

\begin{figure}[H]
\centering
\includegraphics[width=0.46\linewidth]{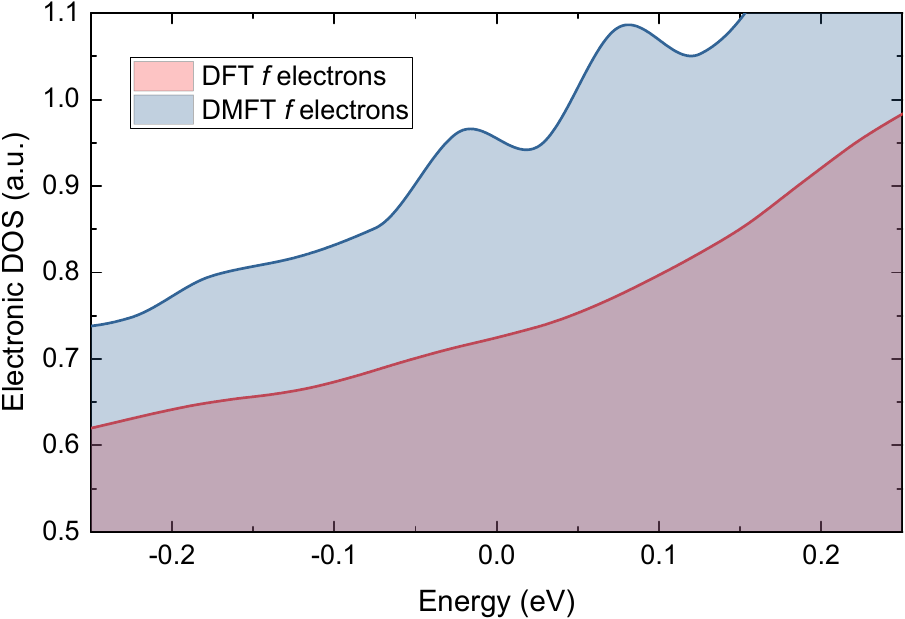}
\caption{Contribution of cerium $f$-electrons to the density of states (DOS) computed with DFT (red) and DFT+DMFT (blue), respectively.}
\label{fig:ceh9-dos}  
\end{figure}

\begin{figure}[H]
\centering
\includegraphics[width=0.46\linewidth]{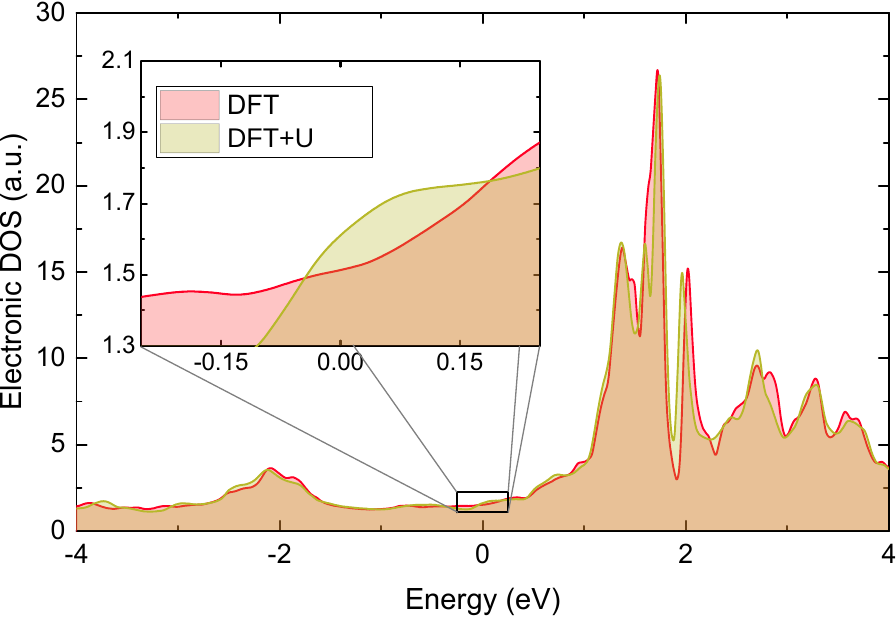}
\caption{Density of states (DOS) of \ce{CeH9} calculated through DFT (red) and DFT+U (yellow).} 
\label{fig:ceh9-dft-band}  
\end{figure}

\section{Section S3: Localized self-energy from DFT+DMFT}
\begin{figure}[H]
\centering
\includegraphics[width=0.89\linewidth]{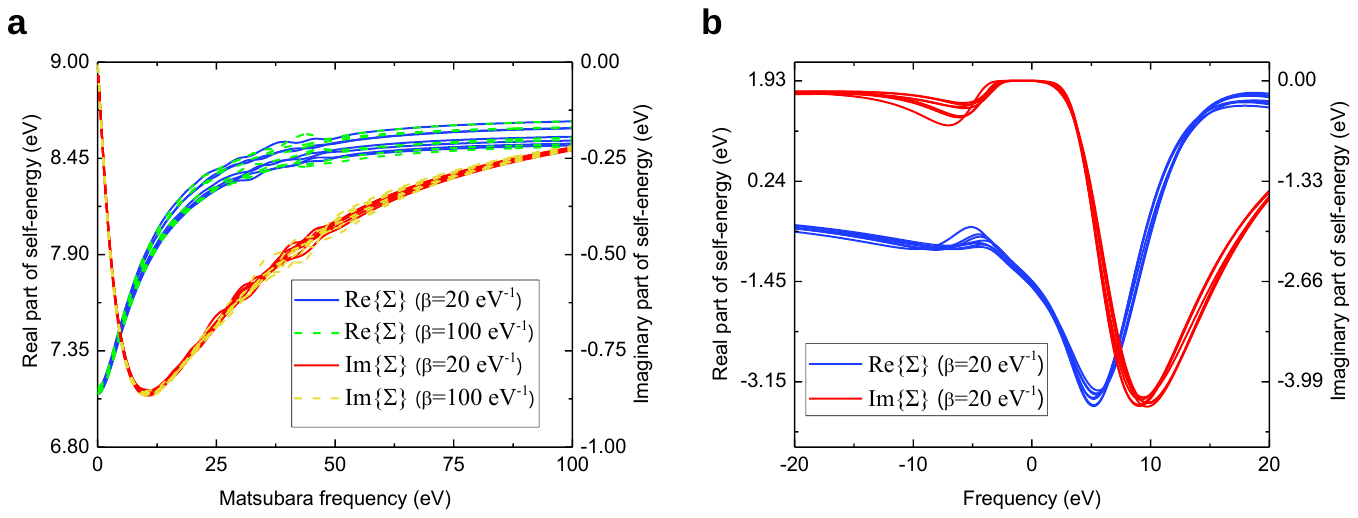}
\caption{Real and imaginary parts of the diagonal terms of the electronic self-energy for the seven $4f$-orbitals, shown along \textbf{a} the imaginary (Matsubara) frequency axis and \textbf{b} the real frequency axis.}
\label{fig:ceh9-self-energy}  
\end{figure}

\section{Section S4: Comparison of DFT and DMFT-renormalized band structures}
\begin{figure}[H]
\centering
\includegraphics[width=0.695\linewidth]{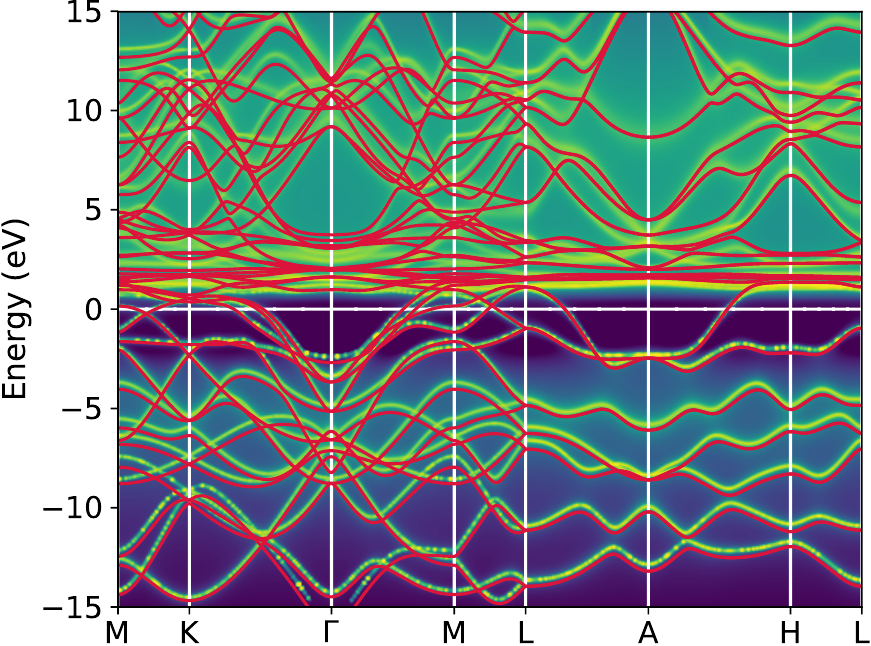}

\caption{Momentum-resolved spectral function $A_\mathbf{k}(\varepsilon)$ of \ce{CeH9} along a high-symmetry path calculated with DFT+DMFT (color map) compared to the DFT bands (red lines).}
\label{fig:ceh9-dmft-spec}  
\end{figure}

\begin{figure}[H]
\centering
\includegraphics[width=0.695\linewidth]{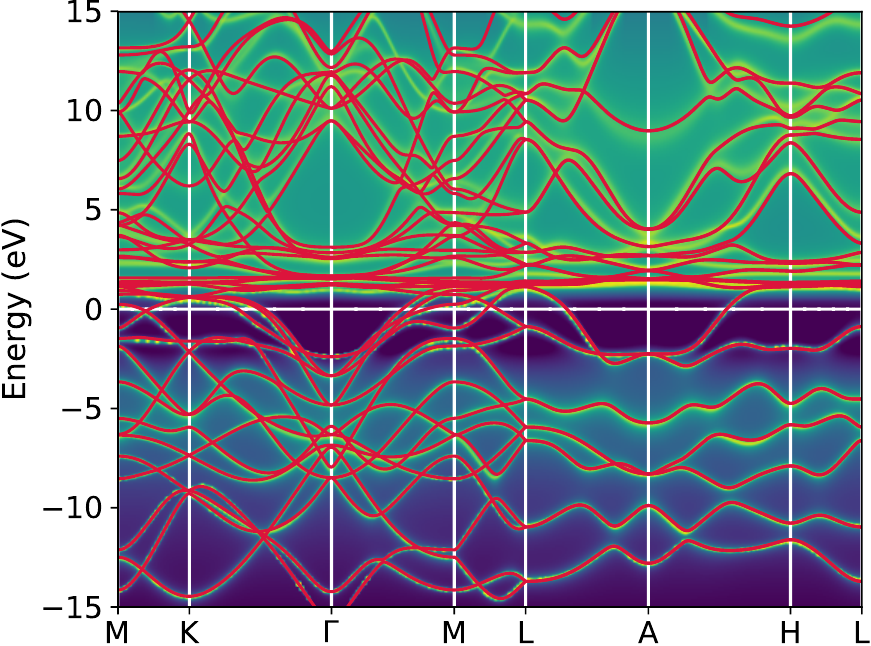}

\caption{Momentum-resolved spectral function $A_\mathbf{k}(\varepsilon)$ of \ce{CeH9} along a high-symmetry path calculated with DFT+DMFT (color map) compared to the DMFT-renormalized bands (red lines).}
\label{fig:ceh9-dmft-spec}  
\end{figure}

\section{Section S5: Nominal occupation}

\begin{figure}[H]
\centering
    \includegraphics[width=0.59\linewidth]{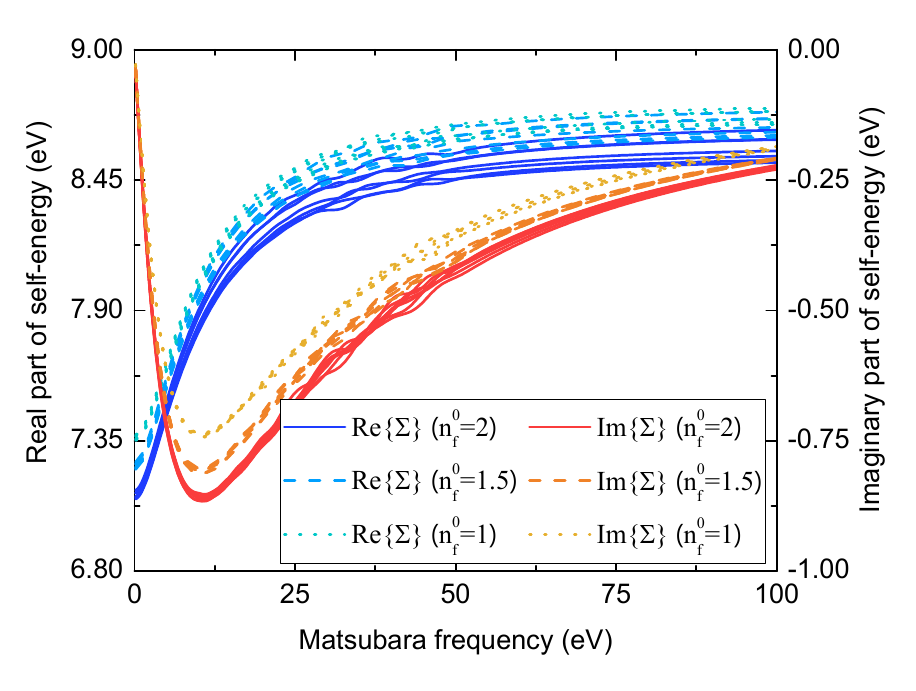}
    \caption{Comparison of the self-energy $\Sigma_L(i\omega)$ computed with the following nominal occupation of $n_f^0$ = 1,  1.5 and 2.}
\end{figure}

\vspace{0.25cm}
\section{Section S6: Local magnetic Susceptibility}

\begin{figure}[H]
\centering
    \includegraphics[width=0.59\linewidth]{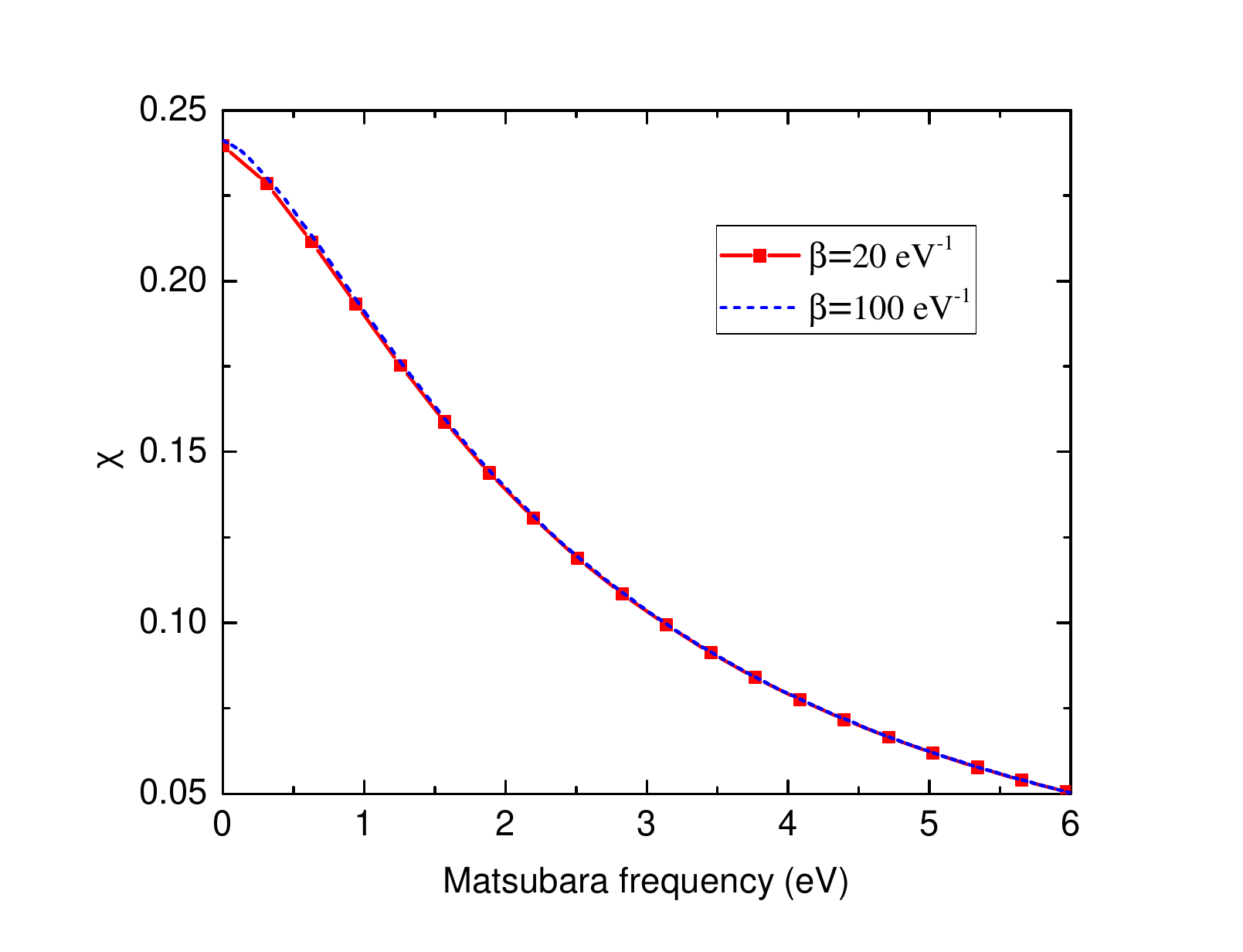}
    \caption{Local magnetic susceptibility of CeH$_9$ as a function of Matsubara frequencies for 580~K and 116~K.}
    \label{fig:ceh9-chi}
\end{figure}

We assess the magnetic properties of CeH$_9$ through the DMFT local magnetic susceptibility, $\chi(i\nu)=\int_0^\beta d\tau \langle S_z(\tau)S_z(0)\rangle e^{i\nu\tau}$, where $S_z(\tau)$ is the $z$-component of the impurity spin operator in the imaginary-time Heisenberg picture. 
As shown in Fig.~\ref{fig:ceh9-chi} for $\beta=20$ eV$^{-1}$ ($T\simeq580$ K) and $\beta=100$ eV$^{-1}$ ($T\simeq116$ K), we find (i) that the susceptibility does not host a contribution $\propto \delta(i\nu=0)$ that would give rise, after Fourier transforming, to a local moment that does not decay in time, and (ii) the susceptibility is essentially temperature independent for the shown range, plausibly ruling out a Kondo-origin of the screening of the $4f$-moment.

\section{Section S7: Comparison of pseudopotential-DFT and all-electron-DFT}
\begin{figure}[H]
\centering
\includegraphics[width=0.45\linewidth]{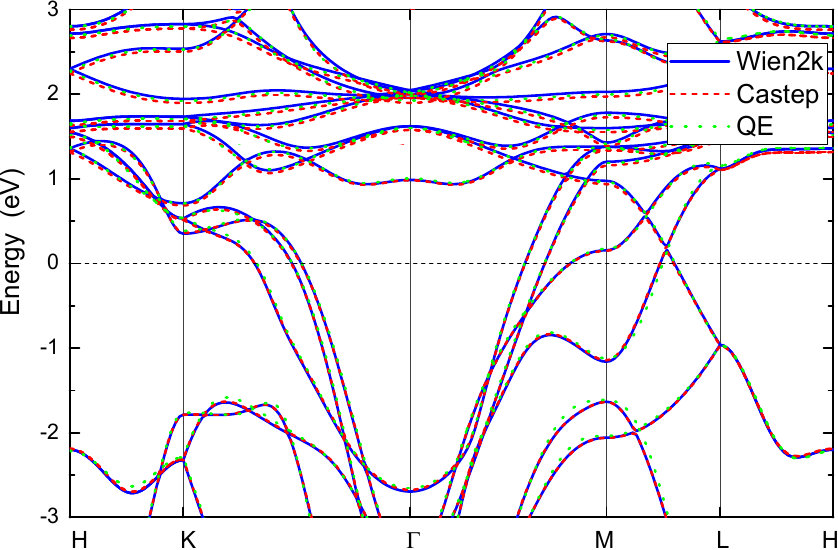}
\caption{Comparison of band structures calculated by \textsc{Wien2k} using the all-electron method and by \textsc{Castep} and \textsc{Quantum ESPRESSO} (QE) using the pseudopotential method.}
\label{fig:xxx}  
\end{figure}

\vspace{1.5cm}
\section{Section S8: Comparison of phonon dispersions }

\begin{figure}[H]
\centering
    \includegraphics[width=0.5\linewidth]
    {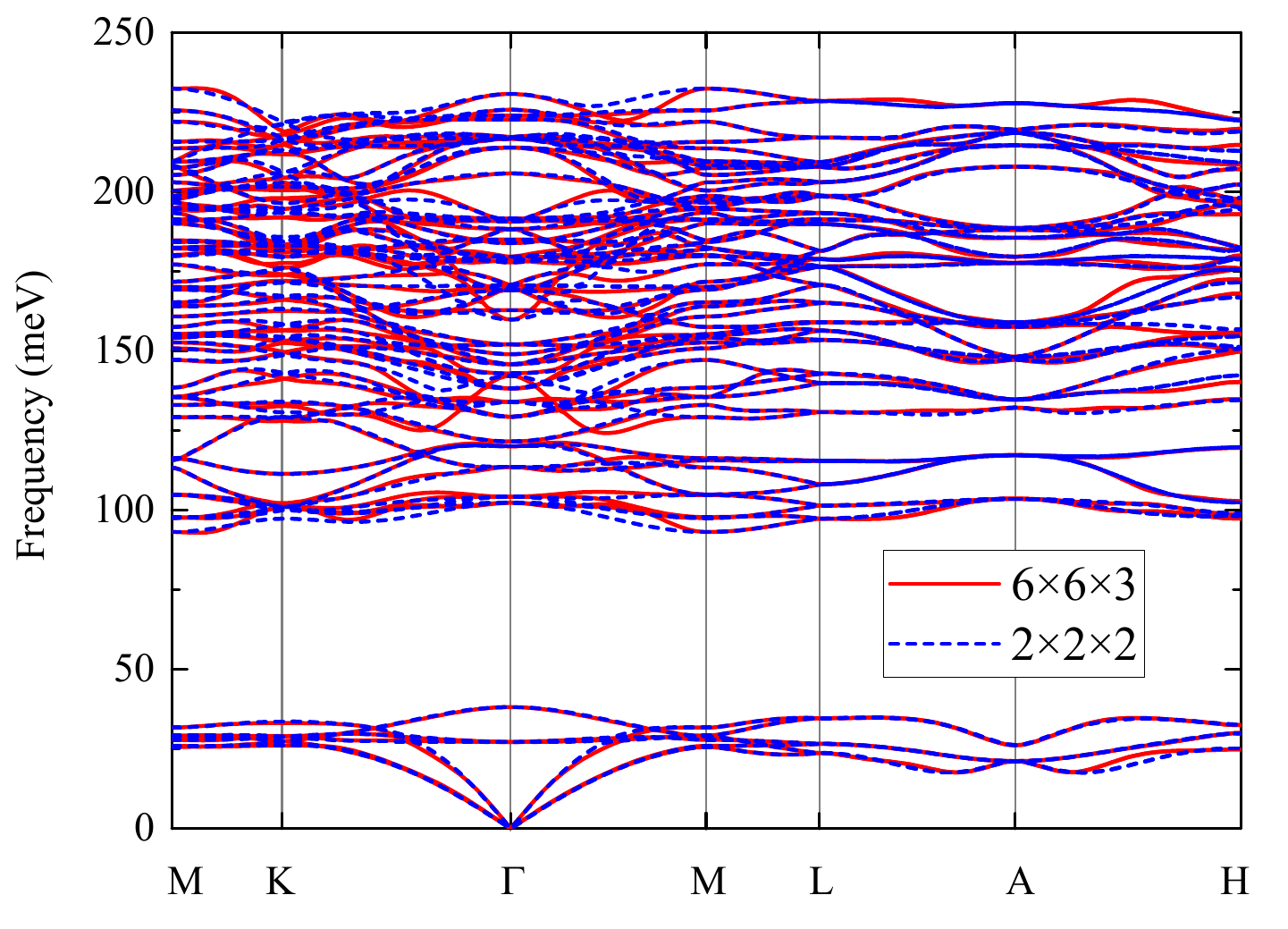}
    \caption{Comparison of phonon dispersions at the DFT level obtained with 2$\times$2$\times$2 and 6$\times$6$\times$3 $\textbf{q}$-point grids}.
\end{figure}

\vspace{1cm}
\section{Section S9: Comparison of self-energy of equilibrium and displaced configurations}

\begin{figure}[H]
\centering
 \quad \quad\quad\includegraphics[width=0.65\linewidth]
    {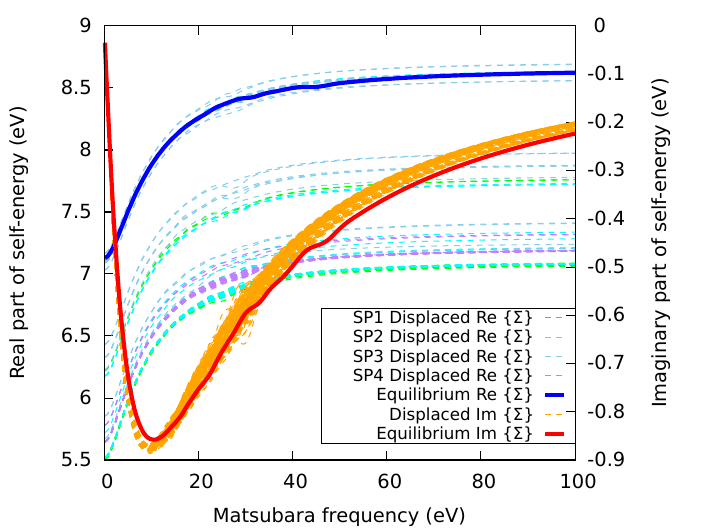}
    \caption{Comparison of self-energy $\Sigma(i\omega)$ of the equilibrium and all displaced configurations, where SP1–SP4 denote four shapes of the supercell used in finite difference calculation.}
\end{figure}

\bibliography{references}